\documentclass[12pt]{revtex4}
\usepackage{ulem}
\usepackage{url}
\usepackage{epsfig}
\usepackage{graphicx,color}
\usepackage[psamsfonts]{amssymb}
\usepackage{amsmath}
\usepackage{indentfirst}

\usepackage[latin1,applemac]{inputenc}

\newcommand{\be}{\begin{equation}}
\newcommand{\ee}{\end{equation}}
\newcommand{\ba}{\begin{eqnarray}}
\newcommand{\ea}{\end{eqnarray}}

\begin{document}
\title{Exploring the limits of safety analysis \\ in complex technological systems}

\author{D. Sornette}
\email{dsornette@ethz.ch}
\affiliation{Department of Management, Technology and Economics,
ETH Zurich, Scheuchzerstrasse 7, CH-8092 Zurich, Switzerland}

\author{T. Maillart}
\email{tmaillart@ethz.ch}
\affiliation{Department of Humanities, Social and Political Sciences,
ETH Zurich, Ramistrasse 101, CH-8092 Zurich, Switzerland}

\author{W. Kr\"oger}
\email{kroeger@mavt.ethz.ch}
\affiliation{Risk Center,
ETH Zurich, Scheuchzerstrasse 7, CH-8092 Zurich, Switzerland}


\date{\today}

\begin{abstract}
\vspace{1cm}
From biotechnology to cyber-risks, 
most extreme technological risks cannot be reliably estimated from historical statistics.
Therefore, engineers resort to predictive methods, such as fault/event trees
in the framework of {\it probabilistic safety assessment} (PSA), which consists in
developing models to identify triggering events, potential accident
scenarios, and estimate their severity 
and frequency. However, even the best safety analysis
struggles to account for evolving risks resulting from inter-connected 
networks and cascade effects.
Taking nuclear risks as an example, the predicted plant-specific
distribution of losses is found to be significantly underestimated
when compared with available empirical records. Using a novel database of 99
events with losses larger than \$50'000 constructed by Sovacool, 
we document a robust power law distribution with tail exponent $\mu \approx 0.7$.
A simple cascade model suggests that the classification of
the different possible safety regimes
is intrinsically unstable in the presence of cascades. Additional
continuous development and validation, making the best use of the experienced realized incidents,
near misses and accidents, is urgently needed to address the existing
known limitations of PSA when aiming at the estimation of total risks.
\end{abstract}

\maketitle

\section{Introduction}

Most innovations are adopted on the premise that the upside gains
largely make up for the downside short- and long-term risks, in particular
through the adoption of safety measures aiming at preventing or mitigating potential losses. 
But innovations are often disruptive and, by essence, break new ground.
This implies that history is a poor guide for risk assessment due to the novelty of the 
technology and the corresponding insufficient statistics.
For highly technical enterprises for which full scale experiments are beyond reach
(such as the Internet and smart grid technologies and associated cyber-risks, 
technological and population growth and climate change, financial innovation and globalization
and the dangers of systemic banking crises),
engineers resort to simulation techniques and scenario-based analyses.

To be concrete, we restrict our discussion to the nuclear industry,
which has been a leader in the development of
state-of-the-art safety analysis, with outstanding efforts aimed at
preventing incidents from becoming major accidents. For this,
{\it probabilistic safety assessment} (PSA) has been
developed as a decision support tool aiming at ensuring a high level
of plant safety limiting the risks of possible release of radioactivity. 
PSA consists in developing fault and event tree models to simulate accidents, their different triggers and 
induced scenarios, their severities as well as their estimated frequency 
\cite{Risk-Safety-book11,krogerchapter}. When performed as an on-going process continuously refined 
using the information of new examples, plant-specific safety analysis has proved very useful for the
implementation of ever better safety barriers, keeping
the advantages of the technology while reducing its undesirable dangers. 
PSA is a well-established discipline with growing applications in support
of rational decision-making involving important technological and societal risks.

The article is organized as follows. Section 2 presents a brief description of
the PSA methodology. Section 3 describes the main predictions of PSA 
with respect to the plant specific ``core damage frequencies'' and 
the ``large early release frequencies'', before comparing them quantitatively 
with the database of losses per event constructed by Sovacool   \cite{sovacool2008},
with special care to demonstrate the robustness of the reported power law
distribution. Section 4 introduces and analyzes a simple conceptual model
of cascades of failures that allows us to rationalize the discrepancy between
predicted and realized losses. Section 5 concludes.

\section{Brief description of probabilistic safety assessment (PSA)}

PSA provides nuclear plant-specific information on risk metrics at three sequential levels
of end states (levels 1-3). 
Level 1 corresponds to  the assessment of the risk of a core damage (core damage frequency or CDF).
Level 2  assesses the size of radioactive releases from the reactor building, in the event of an accident
(large early release frequency or LERF),
in order to develop accident management strategies and identify potential design weaknesses in reactor containment buildings. Core damage frequency (level 1)
and large early release frequency (level 2) are regarded as representative
surrogates to steer design and operations of systems towards achieving
quantitative and qualitative safety goals.
Level 3 evaluates the impact of such releases on the public and the environment
and is used mainly for emergency planning. In the nuclear domain, PSA Levels 1 and 2 are
required to support regulatory activities in most countries (e.g., in the US since 1995 to complement
the deterministic approach within the framework for risk informed regulation).
The PSA methodology has developed into national and international guidelines
(for an history of PSA, see for instance chapter 2.4 in Ref. \cite{Landolt-Bornstein}
and chapter II-b in Ref. \cite{SECY-11-0089}).
International guidelines are presented in Refs. \cite{IAEA2010-1,IAEA2010-2}
and Ref. \cite{SafetySeries50-P-12} (for the later, see
\url{http://www-ns.iaea.org/standards/} for updates and revisions).
This has resulted in widely established and used codes, such as 
SAPHIRE 8 (Systems Analysis Programs for Hands-on Integrated Reliability Evaluations), 
MELCOR (see for instance Ref. \cite{Melcorex}) and MACCS2 
(MELCOR Accident Consequence Code System, Version 2).

 Since many years, PSA is not thought to represent the true risks and to 
 become generalized across sectors, countries and events without
 considerable adjustments. PSA is mainly
 thought of as a platform for technical exchanges on safety matters
 between regulators and the industry, among peers, between designers
 and operators \cite{Molesh}. PSA provides a rigorous and methodical
 way to steer towards measures to achieve safety goals by efficient 
 use of resources and measures. PSA is a fairly well developed mature
 methodology, although diversely implemented and used in different countries, in particular
 in the choice of internal and external events that are included in the analysis.
 
 The basic methodology of PSA is based on logic trees and human
 reliability analysis (HRA) incorporating unintentional failures only, 
 with the uncertainty analysis being restricted to data variation that assume
 traditional (Normal) distributions. Common cause failure (CCF) analysis is included
 by point estimates on fractions of the system, with rare event approximations
 and cut-sets with cut-off values for the quantification of fault trees
 and binning techniques to cope with the large number of sequences.
 PSA uses in general the assumption that a nuclear power plant
 is essentially a closed system, even under catastrophic accident conditions.
There are considerable weaknesses concerning the neglect of cascades
 and the transformation of PSA modeling to consider the nuclear power plan as 
an open system with strong interactions with its
 environment as well as to include the importance of local conditions, in
 particular under severe accident conditions.
 PSA is limited to single units embedded in an ideal environment, in which
 a safety culture is assumed to be ensured, and for which atmospheric transport
 and dispersion dominates (thus neglecting other pathways). This has implications for the conclusions,
 when the nature of the triggering event (earthquakes, flooding / tsunamis)
 affect large areas, and/or other sources of radiation (spent fuel pool) exist, which are
 currently neglected.
There is also limited feed of accident insights back into PSA methodology 
as well as few new approaches and models developed in other sectors and fields
that influence PSA.

 \section{Confronting PSA with reality}
 
 \subsection{Understanding the predictions of PSA}
 
 Probabilistic safety assessment is seldom used for communication but, in reality,
 it is sometimes used to support political decisions, within the public
 debate and by actors other than the nuclear industry such as the insurance industry.
 And this naturally raises the issue of the possible gaps between the occurrence of 
 accidents, public perceptions and the predictions of PSA.
 
 In order to frame correctly the discussion on how to use PSA to estimate
 risk surrogates such as core damage frequency (CDF) and large early
 release frequency (LERF), we stress again that the basic methods of
 PSA levels 1 and 2 use fault/event trees, human
 reliability analysis (HRA) and common cause failure (CCF), to 
 model plant and operator behavior according to principles of
 reductionism in a plant specific way limited to single units.
In this way, PSA models failure of safety systems using
linear semi-dynamic causal chains, with binning techniques
and many approximations allowing quantification.
And severe accident management guidelines  (SAMG) are taken into account
in such framework, as long as they are planned and trained.

PSA level 3 studies are relatively rare, and uses of PSA to estimate risks for the public
are mostly based on levels 1 and 2 studies, which make assumptions about
population behavior, take deterministic (early) and stochastic (late cancer effects)
into account for submersion, inhalation and ingestion exposure pathways. They
are site-specific but extend the calculations of consequences to large distances (800 km)
to capture low dose latent effects.
``Old'' studies include Refs. \cite{wash1400,GermanphaseA,MarshallSizewell83}, which
have been criticized for instance by Marshall \cite{Marshallsecret84} and Speed \cite{Speed-critic}.
Revisions of ``old'' studies include Ref. \cite{Gesellschaft90}, NUREG-1150 \cite{NUREG-1150}
and NEEDS \cite{NEEDS}, the later 
uses generic, simplified tree techniques for EPR at sites in five countries.

Plant specific core damage frequencies (CDFs) for light water 
reactors obtained by PSA vary between $10^{-4}$/reactor-year
and  $10^{-5}$/reactor-year (chapter 2.5.2 in Ref. \cite{Landolt-Bornstein})
and go down to $10^{-6}$/reactor-year  for newly built plants (EPR in Finland
have CDF of $2 \cdot 10^{-6}$/reactor-year) or well back-fitted plant
(such as the KKL plant in Switzerland). These values are accepted by licensing authorities
after thorough review.
Compared with these general CDFs values of $10^{-4}-10^{-5}$/reactor-year,
plant specific LERFs demonstrating containment capabilities are typically
one order of magnitude lower,  in the range $10^{-5}-10^{-6}$/reactor-year.

Concerning fatalities, the above cited studies show maximum consequences of 
about $10^{5}$ latent fatalities at a probability of $10^{-8}$ per reactor-year
or even lower frequency level and steep decrease as improvements are implemented
(see e.g. chapter 2.5.3 of Ref. \cite{Landolt-Bornstein} and Ref. \cite{SECY-11-0089}).
These high consequence numbers are due to an assumed linear dose-risk relationship.

\subsection{Comparison between PSA frequencies and empirical data \label{compPSAemp}}

Over the human history of nuclear power, there has been 3 to 5 
core damage events for a total of about 10'000-15'000 reactor-years, depending
on whether the three core damage of three of the five reactors in the 
Fukushima Dai-ichi nuclear plant in Japan are counted as 1 or 3:
(i) the Three Mile Island accident (TMI) in the US on March 28, 1979,
(ii) the Chernobyl nuclear accident in Ukraine (former Soviet Union) on 26 April 1986 
and (iii) the Fukushima Dai-ichi nuclear disaster following the Tohoku earthquake and tsunami 
on 11 March 2011. 

Using the most conservative figure of $10^{-4}$/reactor-year for CDFs for light water reactor,
one would thus expect typically only $1$ such event over the 10'000-15'000 reactor-years
of the human nuclear plant history. More striking, 
using the most conservative figure for LERF of $10^{-5}$/reactor-year, one would expect
$0.1$ such event over the course of human history and not the 2-4 large releases of
radioactive elements in the environment (here, the Three Mile Island 
accident is not counted
as the release of radioactive gases and radioactive iodine was small).
These discrepancies are worrisome but they deal with small numbers for which
statistics is by definition unreliable and therefore broad conclusions are unreliable.

To advance the discussion, we propose to enrich the statistics by using
complementary metrics, here the complementary cumulative distribution function (CCDF)
(also known as survival distribution) of the losses $S$ per event (accidents and incidents). Here, one
must be clear about the terminology. The US Nuclear Regulatory Commission (NRC)
classifies ``incidents'' are unforeseen events and technical failures 
during normal plant operation with no offsite release of radioactive elements or
severe damage of equipment. In contrast, ``accidents'' usually refer to 
offsite release of radiation and/or damage to plant equipment. In the seven-level system
of the International Nuclear and Radiological Event Scale, 
levels 1-3 are ``incidents'' and level 4-7 are ``accidents.''  The
TMI accident was rated level 5 while both Chernobyl and Fukushima Dai-ichi accidents
are level 7 and these events are the only reported accidents of nuclear plants over the human 
history of civil use of nuclear power. 

Because these accidents concern only
the extreme tail of the distribution, a statistically sound picture can only emerge
by extending the analysis to the body of the distribution. The goal is to populate
the distribution in the region of large and medium size events, to complement
the paucity of statistics on the extreme tail of the distribution, the one associated
with the three major accidents mentioned above. 
For this, we use the database constructed by Sovacool  \cite{sovacool2008},
who extends the definition usually taken by the nuclear industry and regulators
to include incidents that either resulted in the loss of
human life or more than \$50,000 in property damage. This database
includes 99 nuclear events worldwide from 1952 to 2009 that occurred
in different kinds of nuclear facilities.
The database has been constructed by searching historical archives, newspaper
and magazine articles, and press wire reports, leading to a database of 
99 incidents with an estimation of damage they generated, including 
loss of production and property damage \cite{sovacool2008}. 
These nuclear incidents include hydrogen explosion that damage reactor interior, 
fuel rod catching fire and contaminating facility,
partial core meltdown, cooling system malfunctions,
steam generator leaks, fire damage of control
cables leading to disabling core cooling systems,
electrical error causing fire destroying
control lines and main coolant pumps,
mechanical failure during fuel loading causing  
corrosion of reactor and release of radioactivity into the
plant area, failed tube bundles in steam generators,
damaged thermal shield and core barrel support,  damage to recirculation system pipeline, 
instrumentation systems malfunction during startup, 
corroded reactor coolant pumps and shafts leading to reactor shutdowns, and so on.

Figure \ref{figure1} 
shows a one-to-one comparison of the CCDF of damage respectively generated 
from PSA simulations and from the database of the 99 nuclear events 
constructed by Sovacool \cite{sovacool2008}. For the PSA simulations,
we refer to the so-called Farmer curve of the 1950s, superseded by 
the Rasmussen curve of the 1970s \cite{wash1400,Sehgal} for 
100 reactors, which is basically confirmed by others (German Risk Study, figures 8-13 of Ref. \cite{GermanphaseA}) 
or NUREG \cite{NUREG-1150}. It is appreciated by specialists in charge with PSA
that these curves may well underestimate extreme events.
For our purpose to advance the discussion on an evidence based platform,
the PSA curve predicts a rather thin tail while the empirical
distribution of losses is fat-tailed, a power law 
\be
{\rm Pr}({\rm loss} \geq S) = C/S^{\mu}~,
\ee
with an exponent $\mu \approx 0.7$. Concretely, this means that, for
nuclear power events with damage costing more than one billion dollars,
their frequencies are underestimated by about two orders of magnitude. 
Moreover, rather than being associated with just a few extreme cases, the 
existence of a single power law to quantify the distribution of losses over
wildly different magnitudes suggests that the problem has intrinsic structural roots.

\subsection{Robustness of Empirical Distribution}

\begin{figure}[h]
\centerline{\epsfig{figure=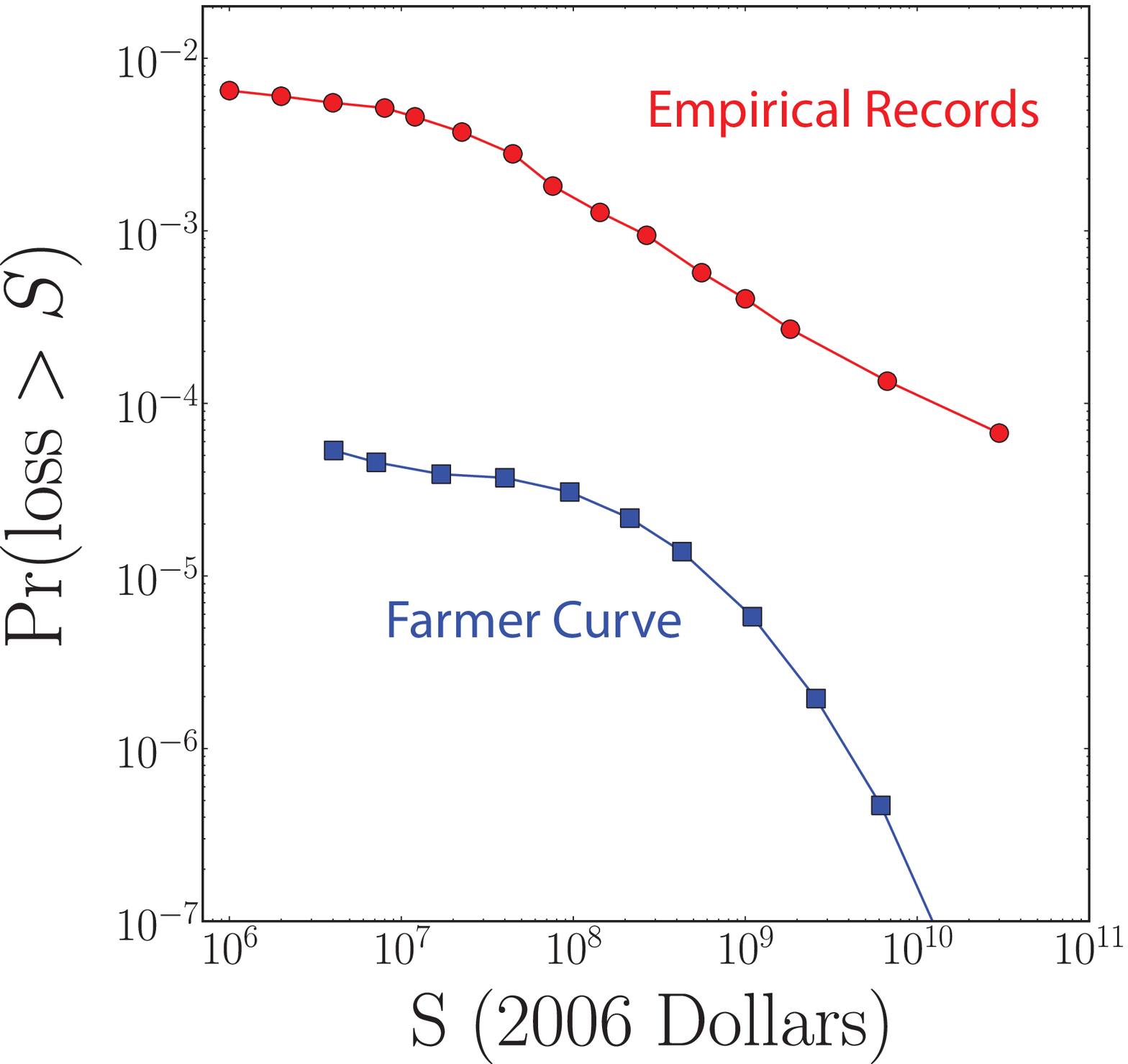,angle=0,width=16cm,scale=1}}
\caption{ (color online) Distribution of losses $S$ 
normalized for $\rm 1~ nuclear~plant \times year$, obtained (i) from probabilistic safety assessment (PSA) and 
(ii) from empirical records \cite{sovacool2008}. We refer to the PSA distribution as the ``Farmer curve'' for
historical reason, even if we use the more recent Rasmussen curve of the 1970s \cite{wash1400,Sehgal} for 
100 reactors, which is basically confirmed by others (German Risk Study, figures 8-13, 14)
or NUREG \cite{NUREG-1150}. Safety analysis largely underestimates the losses due to nuclear incidents. The difference is striking in the tail of the distribution: the distribution obtained from the PSA method vanishes fast beyond \$1 billion damage while empirical records exhibit a power law tail with exponent $\mu = 0.7\pm0.1$ with no apparent cut-off.}
\label{figure1}
\end{figure}

To assess how much to trust the empirical power law distribution of losses reported
in figure  \ref{figure1}, we assess its robustness over time. 
Indeed, an obvious concern with respect to the analysis presented in Figure \ref{figure1} is that 
we are combining events involving different types of nuclear facilities and of 
plants, in particular, with different technologies and generations as well as of varying operational contexts.
But nuclear plants and their safety procedures are continuously updated, resulting in major technological improvements over time \cite{Clery2011}.  This suggests that the distribution of losses should change to reflect that extreme risks are becoming  less and less probable. Indeed, following the TMI (1979) and Chernobyl (1986) accidents, it is a fact that safety has improved. The release of the WASH-1400 Report in the United States has consecrated the adoption of 
PSA compared with more simplistic safety methods, followed by widespread adoption by other countries \cite{wash1400}.  

\begin{figure}[h]
\centerline{\epsfig{figure=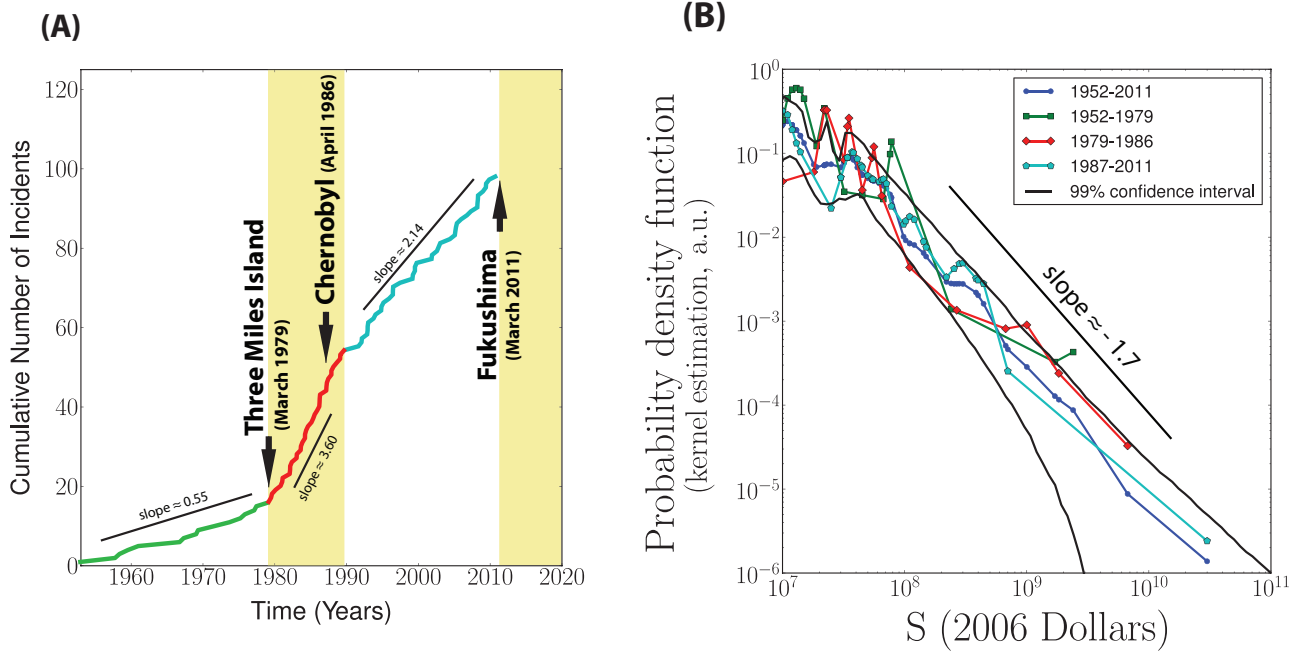,angle=0,width=17cm,scale=1}}
\caption{ (color online) {\bf (A)} Cumulative number of civil nuclear incidents over time since 1957. 
Three regimes can be distinguished: (i) the burgeoning nuclear industry with small
but quickly increasing installed capacity with an average rate of $0.55$ incidents per year;
(ii) from the time of the Three Mile Island accident to the Chernobyl accident, a rate of $3.6$ incidents per year;
(iii) the post-Chernobyl era is characterized by a rate of slightly above $2$ incidents per year. {\bf (B)} Test for stability of the empirical complementary cumulative distribution function (CCDF) over three distinct time intervals of nuclear power industry history and over the whole period, using adaptive kernel density estimators \cite{Cranmer2001}.}
\label{figure2}
\end{figure}

Figure \ref{figure2} tests empirically these points. Panel (A) shows the cumulative number of significant events over time from Ref.~\cite{sovacool2008}. Three regimes can clearly be distinguished. In particular, following the Chernobyl's accident, the rate of incidents has been reduced roughly by 70\%, most likely as a result of additional safety measures worldwide \cite{Clery2011}. Panel (B) of figure \ref{figure2} shows the three empirical distributions of damage in each of these three periods (first period ending at the TMI accident, second period until the Chernobyl accident and third period ending in 2011), together with 
the distribution over the whole period from 1952 to 2011.
It is remarkable that no statistically significant difference can be found, notwithstanding the very different incident rates shown in panel (A) of figure \ref{figure2}. This suggests that, while safety improvements of nuclear plants 
and installations had very positive effects preventing the initiation of events, therefore reducing their occurrence rate, no significant change in the structure of the tail distribution (i.e. of  the relative likelihood of extreme risks compared with small risks) can be observed. This suggests that improved safety procedures and new technology have been quite successful in {\it preventing} incidents (as well as accidents), but mitigation has not significantly improved. This is a paradox, since safety measures should also be designed to minimize the likelihood that, following initiation, incidents worsen and breach one or several of the seven {\it defense-in-depth} safety barriers that protect the infrastructure of a typical nuclear plant. Safety barriers include: (i) prevention of deviation from normal operation, (ii) control of abnormal operation, (iii) control of accidents in design basis, (iv) internal accident management including confinement protection and (v) off-site emergency response 
\cite{insag-10}.

\begin{figure}[h]
\centerline{\epsfig{figure=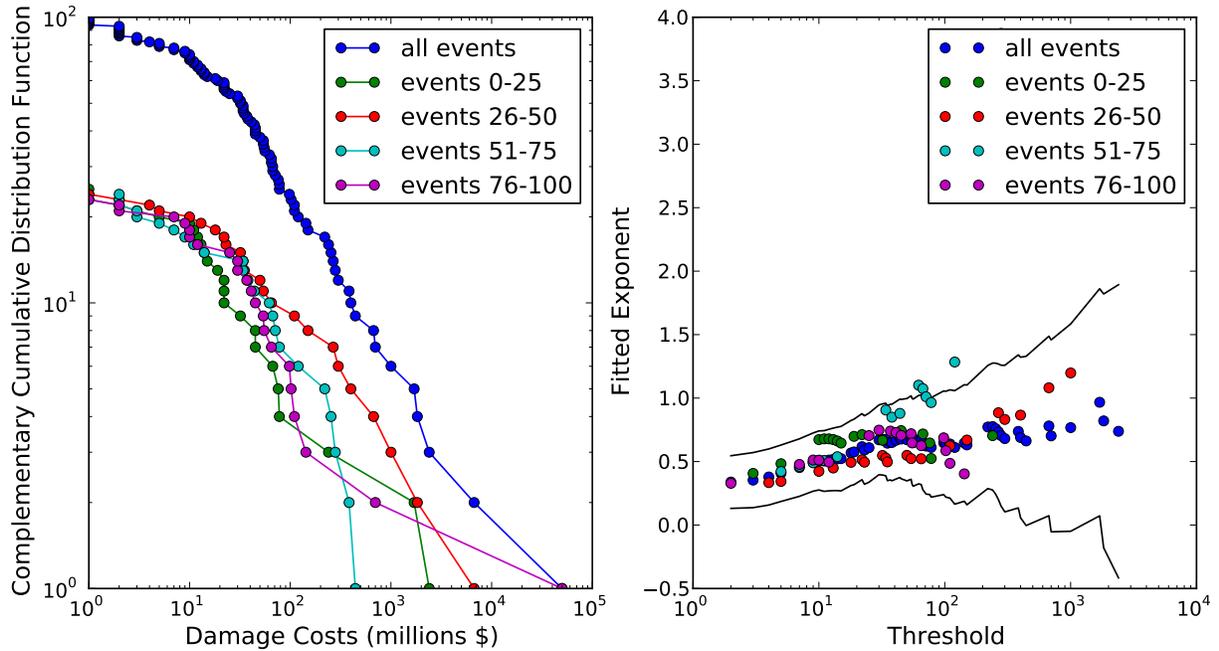,angle=0,width=19cm,scale=1}}
\caption{(color online) Left panel shows the survival distribution functions of loss per nuclear incident, broken 
down in periods of times from 1957 to 2011 with 25 events in each time interval. 
The tails of the distributions for the five time periods have the same power law structure, 
as confirmed by the maximum likelihood estimates (MLE) of the tail exponent $\mu$
for each distribution shown in the right panel. The MLE are calculated for each of the five distributions
by varying the lower threshold in units of million US\$ (abscissa), i.e. by taking into account only the losses
larger than the threshold value given by the abscissa in the right panel. This allows us probing 
the stability of the power law tail. One can observe the exponents clustering around the value $\mu=0.7$
for thresholds larger than \$30 million.
The two continuous lines indicate the 95\% confidence interval for the MLE of the exponent $\mu$.}
\label{figure3}
\end{figure}

As further tests of the reliability and robustness of the power law description, 
the right panel of Figure \ref{figure3} shows the survival distribution 
functions of loss per nuclear incident, broken 
down in periods of times from 1957 to 2011 with 25 events in each time interval. 
The tails of the distributions for the five time periods have the same power law structure, 
as confirmed by the maximum likelihood estimates (MLE) of the tail exponent $\mu$
for each distribution shown in the right panel.

\section{Conceptual model of cascades of failures}

\subsection{Motivations}

In order to rationalize this discrepancy between predicted and realized losses,
we propose a simple conceptual model that embodies the observation that
most important industrial accidents involve cascades with inter-dependent
and mutually amplifying effects \cite{Helbing}. As reviewed above, such cascades
are usually modeled in probabilistic safety assessment by
using fault and event tree techniques \cite{Sehgal}. We suggest
that the discrepancy discussed in subsection \ref{compPSAemp}
may results from the complexity of 
the nonlinear cascades that is likely to obfuscate some critical paths 
that turns out to dominate the generation extreme events. In other words, 
we suggest that simple models may have a role to complement more complex models.
For instance, simple time-dependent statistical models running
on desktops may over-perform large-scale general circulation models requiring super-computers 
for the prediction of climate cycles such as the
Southern Oscillation Index and El Ni\~no event \cite{KeppennGhil}.
The prediction of material failure has been shown to be feasible and 
successful with simple models capturing the essence of the
mechanism of positive feedbacks during the damage process
\cite{Anifrani95,JohSor2000}, an approach that can be extended
to other complex dynamical systems \cite{Sornette02}.
Similarly, complex intermittent dynamics can be quite well represented by simple
Boolean delay equations \cite{GhilBooleandelay}. 
Moreover, it has recently been shown that combining simple and complex models
can improve predictive accuracy and quantify better
predictive uncertainty, with applications to environmental and climate science 
\cite{Balcerek2012}.

In this spirit, we consider a simple model of cascades of failures, in which 
an accident is modeled as a succession of unbroken and broken 
safety barriers with increasing damage. Our goal is to use the simple
conceptual model to translate the discrepancy shown in Figure \ref{figure1}
into technically meaningful insights.

\subsection{Definition and main properties}

Starting from an initiating event generating a damage  $S_{0}$, 
we assume that an event may cascade into a next level with probability $\beta$ 
with associated additional damage $\Lambda_1 \cdot S_{0}$. 
When this occurs, the cascade may continue to the next level, again with 
probability $\beta$ and further additional damage $ \Lambda_2 \Lambda_1 \cdot S_{0}$. 
The probability for the incident to stop after $n$ steps is $P(n) = \beta^{n} (1-\beta)$.
After $n$ steps, the total damage is the sum of the damage at each level:
\be
S_{n} = S_{0} \sum_{k=1}^{n} \Lambda_1 ... \Lambda_k~.
\label{heyhyw}
\ee
Thus, $S_{n}$ is the recurrence solution of the 
Kesten map \cite{Kesten,SornetteCont}:
\be
S_{n} = \Lambda_n S_{n-1} +S_0
\ee
As soon as amplification occurs (technically, 
some of the factors $\Lambda_k$ are larger than $1$), the distribution
of losses is a power law, whose exponent $\mu$ is the function of $\beta$
and of the distribution of the  factors $\Lambda_k$ that solves 
the equation ${\rm E}[\Lambda_k^\mu]=1$ \cite{Kesten,SornetteCont,Sormulpower98}.

The multiplicative mature of the damage represented in expression 
(\ref{heyhyw}) captures a coarse-grained modeling of 
the law of ``proportional growth'' such that future damages are proportional to present
damage, for instance via a branching process. Alternatively, one can view
the multiplicative structure as reflecting a sequence of triggering events, the larger the
existing damage, the larger is the future potential future damage that can be triggered by it.
This proportional growth of damage is not thought to represent faithfully the
details of each damage path but rather embodies a coarse-grained average law.
In a sense, future damage events ``preferentially attachment'', result from or are triggered via
various pathways by existing damage.

In the case where all factors are equal to $\Lambda$, 
this model predicts three possible regimes for the distribution of damage: 
thinner than exponential for $\Lambda < 1$, exponential for $\Lambda = 1$, 
and power law for $\Lambda > 1$ with exponent $\mu = |\ln \beta|/ \ln \Lambda$,
as shown in the next subsection.

\subsection{Solution of the cascade model for homogenous factors $\Lambda_k = \Lambda$}

The following is a detailed study of the possible behaviors of the model, in particular
the different regimes around  the critical point $\Lambda =1$.

Three regimes must be considered:
\begin{enumerate}
\item For $\Lambda < 1$, the distribution is given by
\be
P_{\Lambda <1}(S  \geq s) = (1-\beta) \left(1 - {s \over s_{\rm max}}\right)^c~, ~~~s_{\rm max} := {S_0 \Lambda \over 1-\Lambda}~, ~~c := {\rm{ln}{\beta} 
\over \rm{ln}{\Lambda}} >0~.
\label{trjruyk5i}
\ee
This distribution can be approximated in its central part, away from the maximum possible loss $s_{\rm max}$,
by a Weibull distribution of the form

\be
{\rm Pr}_{\rm }({\rm S} \geq s) \sim e^{-(s/d)^{c}}~.
\label{trjeargquju}
\ee

For $\Lambda \to 1^-$, 
we have $s_{\rm max} \to +\infty$ and, for $s \ll s_{\rm max}$,
expression (\ref{trjruyk5i}) simplifies into a simple exponential function 
\be
P_{\Lambda \to 1^-}(S  \geq s) \sim e^{-|\ln(\beta)| s/S_0}~.
\label{jrtikik}
\ee

\item For $\Lambda = 1$, the distribution of losses  is a simple exponential function since $S_{n} = n S_{0}$ is linear in the number $n$ of
stages and the probability of reaching stage $n$ is the exponential
$P(n) = \beta^{n} (1-\beta)$. Actually, the expression
(\ref{jrtikik}) becomes asymptotical exact as
\be
P_{\Lambda =1}(S  \geq s) = (1-\beta) e^{-|\ln(\beta)| s/S_0}~.
\label{jrtikikwetgt}
\ee

\item For $\Lambda > 1$, the distribution of losses is of the form,

\be
P_{\Lambda >1}(S  \geq s) = \frac{1}{{{(1+ \frac{s}{s*})}^{c}}}~,~~s^{*} := {S_0 \Lambda \over \Lambda -1}~, ~~c := {|\rm{ln}{\beta}|\over \rm{ln}{\Lambda}}~,
\label{jrtisdfjl}
\ee

which develops to a power law distribution of losses of the form
${\rm Pr}({\rm loss} \geq S) = C/S^{\mu}$ with $\mu = c$, when $\Lambda \rightarrow +\infty$.

For $\Lambda \to 1^+$, the tail is still power law and the exponent $\mu$ grows without 
bound if the probability $\beta$ does not converge to $1$.  If both $\Lambda \to 1^+$
and $\beta \to 1^+$ such that $\Lambda = 1-a$, $\beta = 1-\rho a$, with $a \to 0^+$ and $\rho$ 
constant, we have $\mu \to \rho$.
\end{enumerate}

\subsection{Model calibration}

\begin{figure}[h]
\centerline{\epsfig{figure=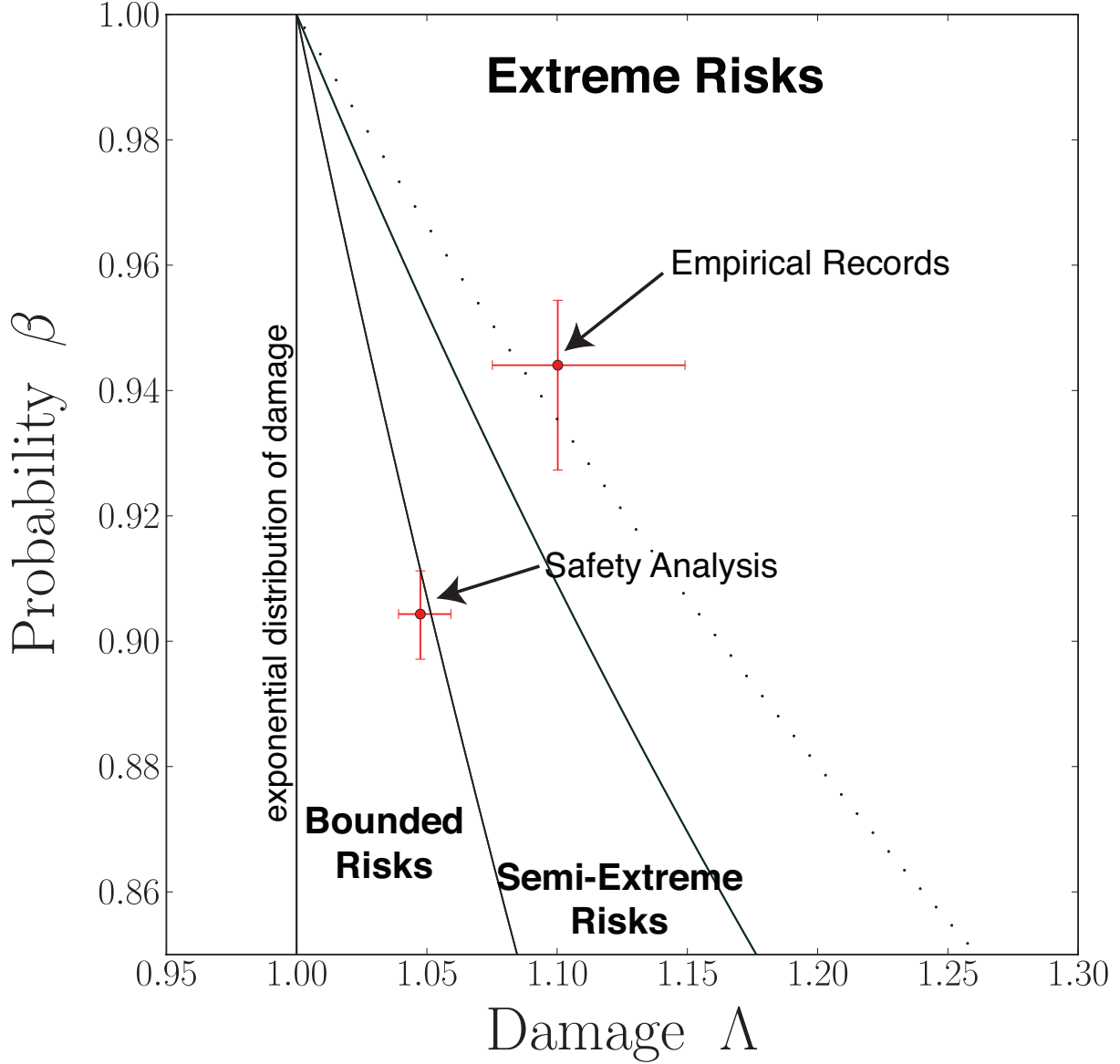,angle=0,width=16cm,scale=1}}
\caption{(color online) Phase diagram showing the three regions of fat-tail risks predicted
by the cascade model: (i) bounded risks with mean and variance defined (exponent $2<\mu$), (ii) semi-extreme risks with only mean defined and variance undefined (exponent $1 <\mu \leq 2$) and (iii) extreme risks with unbounded mean and variance (exponent $\mu \leq 1$). Empirical records clearly identify nuclear accidents as extreme risks (upper right red point with confidence intervals), whereas safety analysis predicts that damage following nuclear incidents is fat-tailed yet bounded (lower left).}
\label{figure4}
\end{figure}

Figure \ref{figure4} presents these different regimes and the corresponding parameters
calibrated to the PSA curves and to the empirical records. 
We calibrate the model by using expression (\ref{jrtisdfjl}) because, 
for both the Farmer curve and empirical records, the solution lies in the region 
$\Lambda > 1$. The calibration consists in finding (i) the probability $\beta$ 
and (ii) the damage factor $\Lambda$ from the {\it ex-ante} 
predictions obtained from the Farmer safety analysis and from the {\it ex-post} historical records. 

\begin{figure}[h]
\centerline{\epsfig{figure=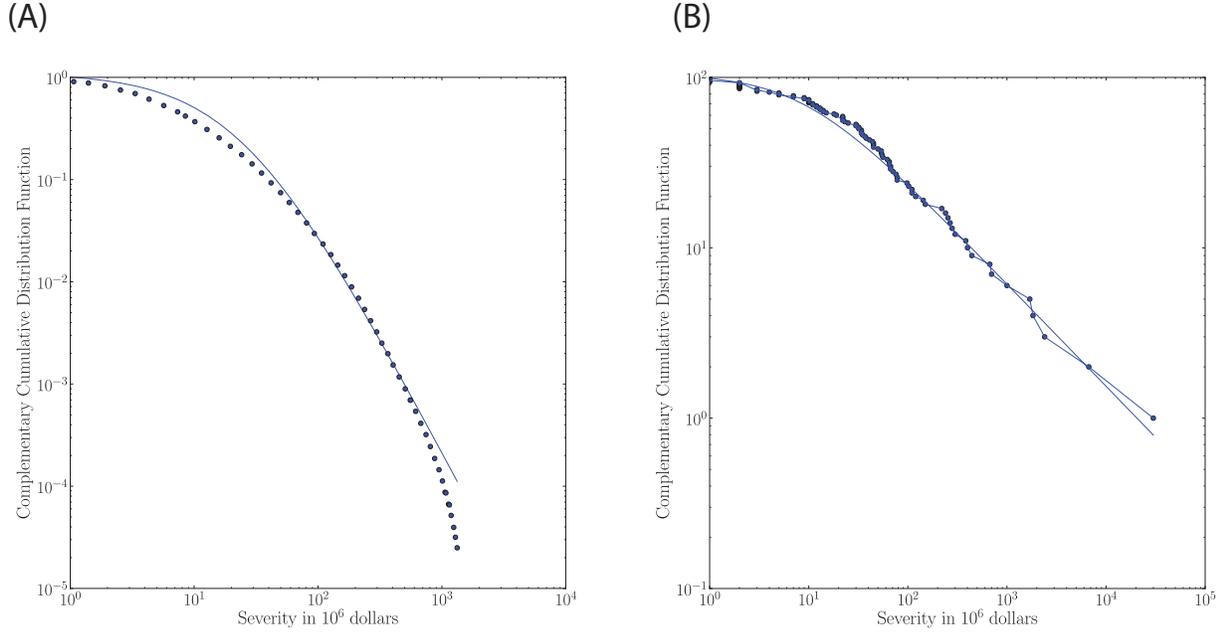,angle=0,width=17cm,scale=1}}
\caption{Calibration of the accident propagation cascade model with
expression (\ref{jrtisdfjl}). 
For the safety analysis (Farmer curve in panel {\bf (A)}), we find that the probability of propagation is $\beta = 0.9$ (lower bound $\beta_{05}=0.9$, upper bound $\beta_{95}=0.91$ at 95\% confidence intervals), and the damage factor is $\Lambda = 1.05$ (lower bound $\Lambda_{05}=1.04$, upper bound $\Lambda_{95}=1.06$, at 95\% confidence interval) for $S_{\rm min}= 10^{7}$. Panel {\bf (B)} shows the best fit of expression (\ref{jrtisdfjl}) to the real data is found for $\beta = 0.95$ (lower bound $\Lambda_{05}=0.94$, upper bound $\Lambda_{95}=0.96$ and $\Lambda = 1.10$ (lower bound $\Lambda_{05}=1.08$, upper bound $\Lambda_{95}=1.15$).}
\label{figure5}
\end{figure}

Both distributions are calibrated first by grid search. From the fitted parameters, one hundred distributions are generated and fitted again by grid search. The final parameters of each distribution are the median values of the bootstrapped distributions.  
The corresponding fits are shown in figure \ref{figure5}
for the Farmer curve (panel A) and for empirical records  (panel B).
We obtain $(\beta_{\rm Farmer} \approx 0.9 ; \Lambda_{\rm Farmer} \approx 1.05)$
compared with $(\beta_{\rm emp.} \approx 0.95 ; \Lambda_{\rm emp.} \approx 1.10)$.
Interpreted within this cascade model, the safety analysis leading to the PSA curve 
attributes roughly a 90\% probability that an incident having reached
a given level of severity may cascade into the next one. To account for the observed
distribution of losses, this number needs to be increased by just 5\% to about a 95\%
probability of cascading. Thus, an underestimation of just 5\% in the probability $\beta$ 
for a cascade to continue and in the additional amplifying factor $\Lambda$ has the effect
of leading to a significant underestimation of the distribution of losses, in particular for large events. 
The origin of this sensitivity stems from the proximity of $\beta$ to 
the critical value $1$, likely due to optimization as occurs in almost 
any human enterprise associated with the design,
modeling, planning and operating of many complex real-world  problems 
\cite{SorSOCself,CarlsonDoyle,Sornettebook04}.

\section{Conclusions}

Development of probabilistic safety assessment (PSA) at level 3,
which evaluates the impact of unintended radioactive releases on the public and the environment,
has been slowed down and only a few studies exist including a) modifications to enhance
operational performance, safety and security enforcing measures, b) improved
understanding and modeling of several accident phenomena and potential
event sequences, c) advances in PSA technology and other, including simulation 
techniques and use of huge data. For example, the US NRC
has not sponsored development of PSA level 3 since 1990 \cite{SECY-11-0089}.

In this article, we have presented a structured analysis based on empirical evidence that 
suggests a poor track record of PSA to reflect adequately the true risks involved
in the nuclear industry. In particular, PSA is currently not adapted to identify
correctly the likelihood of large catastrophes.   
The recognition obtained from our study of (i) extreme heavy-tailed losses and (ii)
the critical nature of cascades during incidents and accidents 
calls for the integration of the observations and statistics over a large heterogeneous
set of incidents into the PSA framework, a procedure that is at present not implemented 
and seems still far from being considered. If the estimation of total 
risk is the aim, there is an urgent need to address
limitations (despite the acknowledged achievements) and strive to 
overcome them. Safety analysis should be a never ending process constantly building on past experience,
including the evolving full distribution of losses, for the
development and the implementation of ever improved measures based on 
model update \cite{SorKammval}.

A word of caution should be added, as we use a non-standard database, put together by
Sovacool  \cite{sovacool2008}, which represents a best effort, but cannot be considered 
as sufficiently reliable to draw policy recommendations, for instance. 
However, our method to measure the gap between safety analysis and the size of events at 
the coarse grained level is a step towards a systematic and continuous comparison 
between safety analysis and the actual size of events (e.g. in financial loss), at the scale of one facility, 
or at the scale of a utility (e.g. EDF). Applying our approach to smaller scale studies 
would also help overcome the problem of the heterogeneity of nuclear power production 
technologies, which cannot be addressed currently due to the paucity of our data. 
A utility company which would have gathered data on both (i) distributions generated by PSA, 
and (ii) loss records, would be in a position to test and refine PSA by using our method.

The method proposed here applies beyond the nuclear industry to many other
human innovations and technological developments for which
history is a poor guide due to the novelty of the 
technology and the corresponding lack of history leading to insufficient 
statistics, in particular for extreme events. 
Among such difficult unsolved questions, one can cite
the impact of cell phones on brain cancer, genetically modified foods and  
health risks, antibiotic use in cattle and resistant bacterias, biotechnological medicine
and health risks, nanotechnology and health risks, Internet and grid technology
versus cyber-risks, unequal distribution of gains from capital markets and 
the (in-)stability of societies, virtual technology and brain development
in children and teenagers, human population and technological development
and climate change, financial globalization
and systemic banking meltdown, nuclear energy and chemical
industry versus the risks of release of long-lived toxic elements in the biosphere, and so on.

\pagebreak

\pagebreak

\pagebreak

\pagebreak

\pagebreak

\end{document}